\newcommand{\TypeOfDoc}{TR} 
\documentclass[letterpaper, 10 pt, conference]{TemplateFiles/ieeeconf}  

\IEEEoverridecommandlockouts                              

\usepackage{amsmath} 
\usepackage{amssymb}  
\usepackage[ruled,vlined,linesnumbered]{TemplateFiles/algorithm2e}
\usepackage[usenames]{color}
\usepackage[svgnames]{xcolor}
\usepackage{mathrsfs}
\usepackage{graphicx}
\usepackage{amsfonts}
\usepackage{array}
\usepackage{flafter}
\usepackage{subfigure}
\usepackage{verbatim}
\usepackage{psfrag}
\usepackage{umoline}
\usepackage{hyperref}
\usepackage{epstopdf}

\hypersetup{%
    pdfborder = {0 0 0}
}

\graphicspath{{./figs/}}

\def\ds/{Dec-POSMDP} 
\def\dss/{Dec-POSMDPs}

\newcommand{\sXX}[1]{\color{blue}S: (#1) \color{black}}

\newcommand{\cXX}[1]{\color{green}C: #1 \color{black}}

\newtheorem{Def}{Definition}

\newcommand{\iftr}[1]{\ifthenelse{\equal{\TypeOfDoc}{TR}}{\color{black}#1\color{black}\xspace}{}} 
\newcommand{\ifco}[1]{\ifthenelse{\equal{\TypeOfDoc}{conference}}{\color{black}#1\color{black}\xspace}{}} 

\title{\LARGE \bf
Decentralized Control of Partially Observable Markov Decision Processes using Belief Space Macro-actions}

\author{Shayegan Omidshafiei, Ali-akbar Agha-mohammadi, Christopher Amato, Jonathan P. How
\thanks{Omidshafiei, Agha, and How are with the Laboratory for Information and Decision Systems (LIDS), MIT. Amato is with the Computer Science and Artificial Intelligence Laboratory (CSAIL), MIT, Cambridge, MA.
{\tt\small \{shayegan,aliagha,camato,jhow\}@mit.edu}}%
}

\newcounter{Lcount}
\newenvironment{flushenum}{\begin{list}{$\bullet$}{\usecounter{Lcount} \leftmargin=0em \itemindent=5pt}}{\end{list}}

\renewcommand{\baselinestretch}{0.995}

\begin{document}

\maketitle
\thispagestyle{empty}
\pagestyle{empty}

\begin{abstract}
The focus of this paper is on solving multi-robot planning problems in continuous spaces with partial observability. 
Decentralized partially observable Markov decision processes (Dec-POMDPs) are general models for multi-robot coordination problems, but representing and solving Dec-POMDPs is often intractable for large problems. 
To allow for a high-level representation that is natural for multi-robot problems and scalable to large discrete and continuous problems, this paper extends 
the Dec-POMDP model to the decentralized partially observable semi-Markov decision process (Dec-POSMDP). The Dec-POSMDP formulation allows asynchronous decision-making by the robots, which is crucial in multi-robot domains. We also present an algorithm for solving this Dec-POSMDP which is much more scalable than previous methods since it can incorporate closed-loop belief space macro-actions in planning. These macro-actions are automatically constructed to produce robust solutions. 
The proposed method's performance is evaluated on a complex multi-robot package delivery problem under uncertainty, showing that our approach can naturally represent multi-robot problems and provide high-quality solutions for large-scale problems.

\end{abstract}



\section{Introduction} \label{subsec:intro}
Many real-world multi-robot coordination problems operate in continuous spaces where robots possess partial and noisy sensors. 
In addition, asynchronous decision-making is often needed due to stochastic action effects and the lack of perfect communication. 
The combination of these factors makes control very difficult. 
Ideally, high-quality controllers for each robot would be automatically generated based on a high-level domain specification. In this paper, we present such a method for both formally representing multi-robot coordination problems and automatically generating local planners based on the specification. 
While these local planners can be a set of hand-coded controllers, we also present an algorithm for automatically generating controllers that can then be sequenced to solve the problem. 
The result is a principled method for coordination in probabilistic multi-robot domains.


\begin{figure}[t!]
  \centering
 \includegraphics[width=0.5\textwidth]{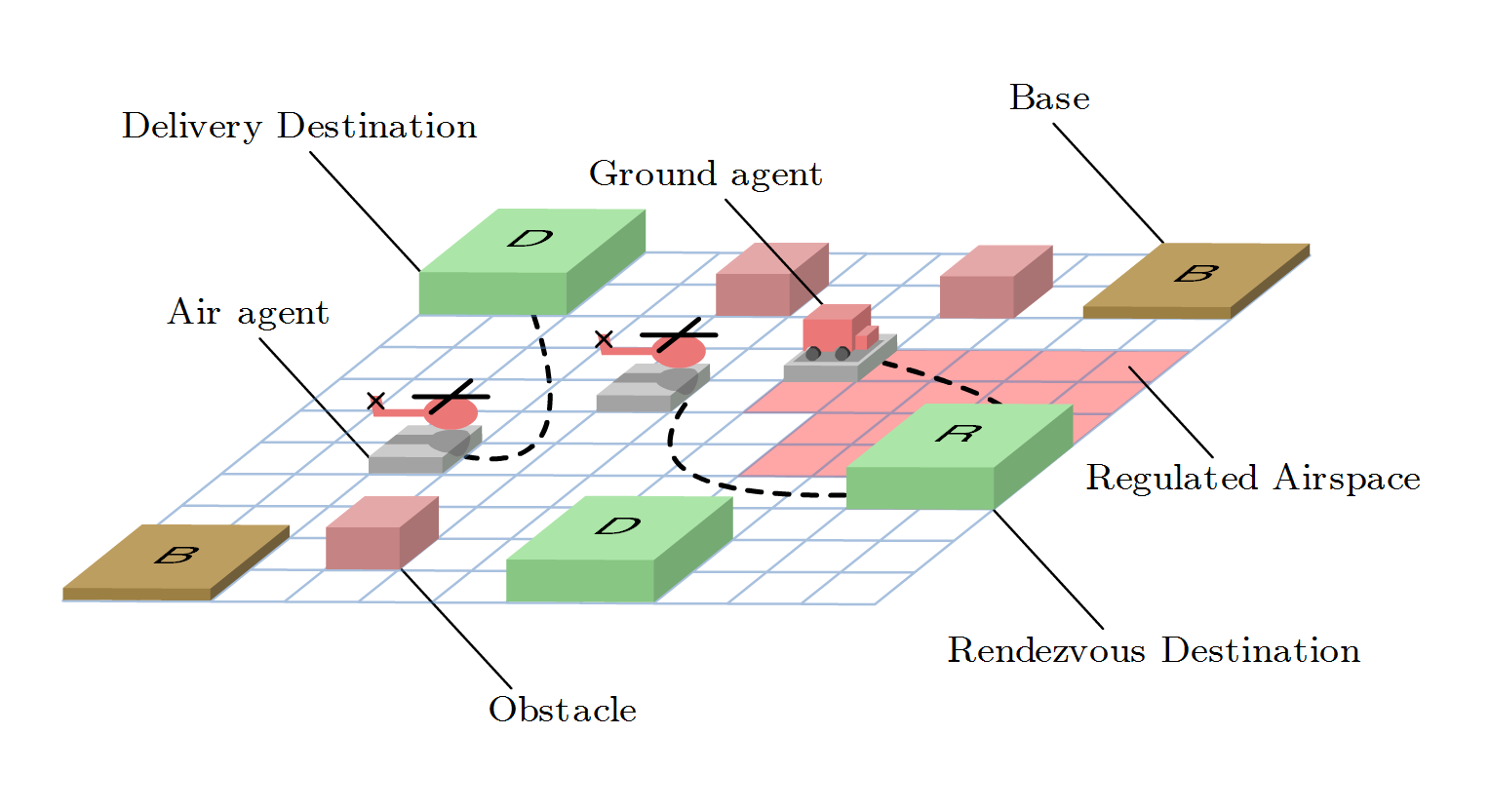}
  \vspace{-25pt}
  \caption{Package delivery domain with key elements labeled.}
 \label{fig:Experiment Scenario} 
\end{figure}



The most general representation of the multi-robot coordination problem is the decentralized partially observable Markov decision process (Dec-POMDP)  \cite{Bernstein02}.
Dec-POMDPs have a broad set of applications including networking problems, multi-robot exploration, and surveillance \cite{Bernstein01,Emery05,Winstein13,Ure13_BookChapter}.
Unfortunately, current Dec-POMDP solution methods are limited to small discrete domains and require synchronized decision-making. 
 This paper extends some promising recent work on incorporating macro-actions, temporally extended actions,  \cite{ICRA15MacDec,AAMAS14AKK}
 to solve continuous and large-scale problems which were infeasible for previous methods.


Macro-actions (MAs) 
have provided increased scalability in single agent MDPs \cite{Sutton99} and POMDPs \cite{Ali14-IJRR,He11JAIR}, but they are nontrivial to extend to multi-agent settings. 
Some of the challenges in extending MAs to decentralized settings are:

\begin{flushenum}
\item In the decentralized setting, synchronized decision-making is problematic (or even impossible) as some robots must remain idle while others finish their actions. The resulting solution quality would be poor (or not implementable), resulting in the need for MAs that can be chosen asynchronously by the robots (an issue that has not been considered in the single agent literature). 

\item Incorporating principled asynchronous MA selection is a challenge, because 
it is not clear how to choose optimal MAs for one robot while other robots are still executing. Hence, a novel formal framework is needed to represent Dec-POMDPs with asynchronous decision-making and MAs that may last varying amounts of time.

\item Designing these variable-time MAs also requires characterizing the stopping time and probability of terminating at every goal state of the MAs. Novel methods are needed that can provide this characterization. 
%
%

\end{flushenum}



MA-based Dec-POMDPs alleviate the above problems by no longer attempting to solve for a policy at the primitive action level, but instead considering temporally-extended actions, or MAs. This also addresses scalability issues, as the size of the action space is considerably reduced.

In this paper, we extend the Dec-POMDP to the decentralized partially observable \emph{semi-}Markov decision process (\ds/) model, which formalizes the use of closed-loop MAs. The \ds/ represents the theoretical basis for asynchronous decision-making in Dec-POMDPs. We also automatically design MAs using  graph-based planning techniques. 
 The resulting MAs are closed-loop and the completion time and success probability can be characterized analytically, allowing them to be directly integrated into the \ds/ framework. As a result, our framework can generate efficient decentralized plans which take advantage of estimated completion times to permit asynchronous decision-making. 
The proposed \ds/ framework enables solutions for large domains (in terms of state\slash action\slash observation space) with long horizons, which are otherwise computationally intractable to solve. We leverage the Dec-POSMDP framework and design an efficient discrete search algorithm for solving it, and demonstrate the performance of the method for the complex problem of multi-robot package delivery under uncertainty (Fig. \ref{fig:Experiment Scenario}).

\begin{figure}[t!]
  \centering
 \includegraphics[width=3 in]{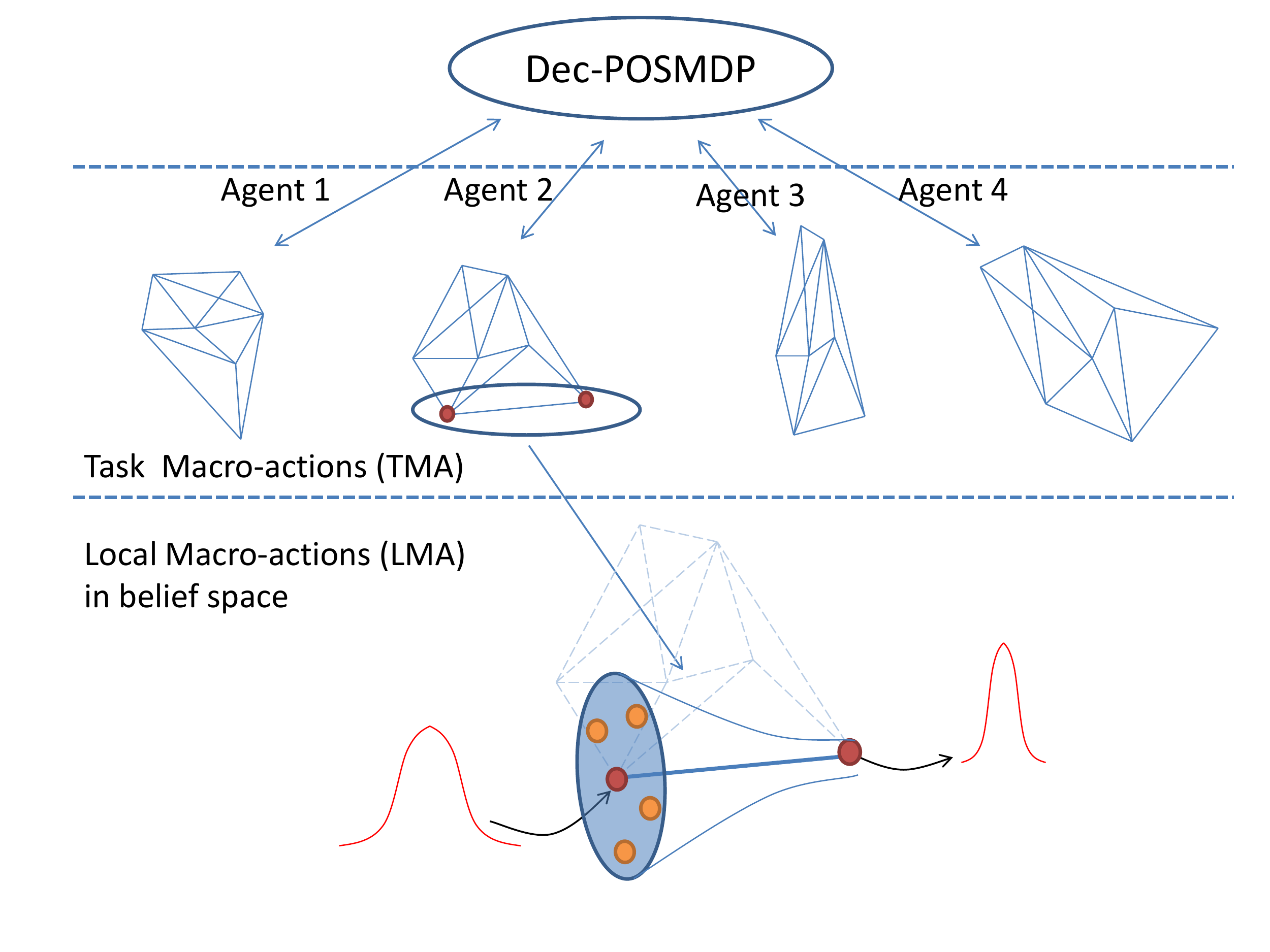}
 \vspace{-5pt}
 \caption{Hierarchy of the proposed planner. In the highest level, a decentralized planner assigns a TMA to each robot. Each TMA encompasses a specific task (e.g. picking up a package). Each TMA in turn is constructed as a set of local macro-actions (LMAs). Each LMA (the lower layer) is a feedback controller that acts as a funnel. LMAs funnel a large set of beliefs to a small set of beliefs (termination belief of the LMA).
 }
 \label{fig:hierarchy} 
\end{figure}




\section{Problem Statement} \label{subsec:problem}



A Dec-POMDP \cite{Bernstein02} is a sequential decision-making problem where multiple agents (e.g., robots) operate under uncertainty based on different streams of observations. At each step, every agent chooses an action (in parallel) based purely on its local observations, resulting in an immediate reward and an observation for each individual agent
 based on  stochastic (Markovian) models over continuous states, actions, and observation spaces. 



We define a notation aimed at reducing ambiguities when discussing single agents and multi-agent teams. A generic parameter $p$ related to the $ i $-th agent is noted as $p^{(i)}$, whereas a joint parameter for a team of $n$ agents is noted as $\bar{p} = (p^{(1)}, p^{(2)}, \cdots, p^{(n)})$. Environment parameters or those referring to graphs are indicated without parentheses, for instance $p^i$ refers to a parameter of a graph node, and $p^{ij}$ to a parameter of a graph edge.

Formally, the Dec-POMDP problem we consider in this paper is described by the following elements:
\begin{itemize}
\item $\mathbb{I}=\{1,2,\cdots,n\}$ is a finite set of agents' indices.
\item $\bar{\mathbb{S}}$ is a continuous set of joint states. Joint state space can be factored as $ \bar{\mathbb{S}}=\bar{\mathbb{X}}\times\mathbb{X}^{e} $ where $ \mathbb{X}^{e} $ denotes the environmental state and $\bar{\mathbb{X}}=\times_i \mathbb{X}^{(i)}$ is the joint state space of robots, with $ \mathbb{X}^{(i)} $ being the state space of the $ i $-th agent. $ \mathbb{X}^{(i)} $ is a continuous space. We assume $ \mathbb{X}^{e} $ is a finite set.
\item  $\bar{\mathbb{U}}$ is a continuous set of joint actions, which can be factored as $\mathbb{U}=\times_i \mathbb{U}^{(i)}$, where $\mathbb{U}^{(i)}$ is the set of actions for the $ i $-th agent.
\item  State transition probability density function is denoted as $ p(\bar{s}'|\bar{s},\bar{u}) $, that specifies the probability density of transitioning from state $\bar{s} \in \bar{\mathbb{S}}$ to $\bar{s}' \in \bar{\mathbb{S}}$ when the actions $\bar{u}\in \bar{\mathbb{U}}$ are taken by the agents.
\item  $\bar{R}$ is a reward function: $\bar{R}: \bar{\mathbb{S}} \times \bar{\mathbb{U}} \to \mathbb{R}$, the immediate reward for being in joint state $\bar{x} \in \bar{\mathbb{X}}$ and taking the joint action $\bar{u} \in \bar{\mathbb{U}}$. 
\item $\bar{ \Omega} $ is a continuous set of observations obtained by all agents. It can be factored as $ \bar{\Omega} = \bar{\mathbb{Z}}\times\bar{\mathbb{Z}}^{e} $, where $\bar{\mathbb{Z}}=\times_i\mathbb{Z}^{(i)}$ and  $\bar{\mathbb{Z}}^{e}=\times_i\mathbb{Z}^{e(i)}$. The set $ \mathbb{Z}^{(i)}\times\mathbb{Z}^{e(i)} $ is the set of observations obtained by the $ i $-th agent. $ \mathbb{Z}^{e(i)} $ is the observation signal that is a function of the the environmental state $ x^{e} \in \mathbb{X}^{e} $. We assume the set of environmental observations $ \mathbb{Z}^{e(i)} $ is a finite set for any agent $ i $.
\item  Observation probability density function $h(\bar{o}|\bar{s},\bar{u})$ encodes the
probability density of  seeing  observations $\bar{o} \in \bar{\Omega}$ given 
joint action $\bar{u} \in \bar{\mathbb{U}}$ taken and resulted in joint state $\bar{s} \in \bar{\mathbb{S}}$.
\end{itemize}
Note that a general Dec-POMDP need not have a factored state space such as the one given here. 

The solution of a Dec-POMDP is a collection of decentralized policies $ \bar{\eta}=(\eta^{(1)},\eta^{(2)},\cdots,\eta^{(n)}) $. Because (in general) each agent does not have access to the observations of other agents, each policy $ \eta^{(i)} $ maps the individual data history (obtained observations and taken actions) of the $ i $-th agent into its next action: $ u^{i}_{t}=\eta^{(i)}(H^{(i)}_{t}) $, where $ H^{(i)}_{t} = \{o^{(i)}_{1},u^{(i)}_{1},o^{(i)}_{2},u^{(i)}_{2},\cdots,o^{(i)}_{t-1},u^{(i)}_{t-1},o^{(i)}_{t}\} $.

According to above definition, we can define the value associated with a given policy $ \bar{\eta} $ starting from an initial joint state distribution $ \bar{b} $:
\begin{align}
V^{\bar{\eta}}(\bar{b})=\mathbb{E} \left[ \sum_{t=0}^{\infty} \gamma^t\bar{R}(\bar{s}_{t},\bar{u}_{t}) | \bar{\eta},p(\bar{s}_{0})=\bar{b} \right]
\end{align}
Then, a solution to a Dec-POMDP formally can be defined as the optimal policy:
\vspace{-8pt}
\begin{align}\label{eq:dec-pomdp}
\bar{\eta}^{*}=\arg\max_{\bar{\eta}}V^{\bar{\eta}}
\end{align}


The Dec-POMDP problem stated in \eqref{eq:dec-pomdp} is undecidable over continuous spaces without additional assumptions. 
Recent work has extended the Dec-POMDP model to incorporate macro-actions which can be executed in an asynchronous manner \cite{AAMAS14AKK}. In planning with MAs, decision making occurs in a two layer manner (see Fig. \ref{fig:hierarchy}). A higher-level policy will return a MA for each agent and the selected MA will return a primitive action to be executed.
This approach is an extension of the \emph{options framework}~\cite{Sutton99} to multi-agent domains while dealing with the lack of synchronization between agents.  
The options framework is a formal model of MAs \cite{Sutton99} that has been very successful in aiding representation and solutions in single robot domains \cite{Kober13}. 
Unfortunately, this method requires a full, discrete model of the system (including macro-action policies of all agents, sensors and dynamics). 

As an alternative, we propose a \ds/ model which only requires a high-level model of the problem. 
The \ds/ provides a high-level discrete planning formalism which can be defined on top of continuous spaces. As such, we can approximate the continuous multi-robot coordination problems with a tractable Dec-POSMDP formulation.  
Before defining our more general \ds/ model, we first discuss the form of MAs that allow for efficient planning within our framework. 

\section{Hierarchical Graph-based Macro-actions} \label{sec:FIRM}
This section introduces a mechanism to generate complex MAs based on a graph of lower-level simpler MAs, which is a key point in solving Dec-POSMDPs without explicitly computing success probabilities, times, and rewards of MAs in the decentralized planning level. We refer to the generated complex MAs as Task MAs (TMAs).

\ifco{
As we discuss in the next section, for a seamless incorporation of MAs into \ds/ planning, we need to design closed-loop MAs, which is a challenge in partially-observable settings. 
In this paper, we utilize information roadmaps \cite{Ali14-IJRR} as a substrate to efficiently and robustly generate such TMAs. 
We start by discussing the structure of feedback controllers in partially observable domains.
}

\iftr{
 To clarify the concepts, before describing task macro-actions, we distinguish between open-loop and closed-loop MAs. In general,  MAs refer to temporally-extended actions \cite{Theocharous04}. An $ l $-step long open-loop MA is a sequence of pre-defined actions such as by $u_{0:l}=\{u_{0},u_{1},\cdots,u_{l}\}$. However, a closed-loop MA is policy $\pi(\cdot)$, i.e., a mapping from histories to actions. As we discuss in the next section, for a seamless incorporation of MAs in Dec-POMDP planning, we need to design closed-loop MAs, which is a challenge in partially-observable settings.


Generating a closed-loop MA that accomplishes a single-agent task such as picking up an object, delivering a package, opening a door, and so on, itself requires solving a POMDP problem. In this paper, we utilize information roadmaps \cite{Ali14-IJRR} as a substrate to generate such task MAs. 
We start by discussing the structure of feedback controllers in partially-observable domains.
}

A macro-action $ \pi^{(i)} $ for the $ i $-th agent maps the histories $ H^{\pi^{(i)}} $ of actions and observations that have occurred to actions. Note that the environmental observations are only obtained when the MA terminates. We compress this history into a belief $ b^{(i)}=p(x^{(i)}|H^{\pi^{(i)}}) $, with joint belief for the team denoted by $\bar{b} = (b^{(1)}, b^{(2)}, \cdots, b^{(n)})$. It is well known \cite{Kumar-book-86} that making decisions based on belief $ b^{(i)} $ is equivalent to making decisions based on the history $ H^{\pi^{(i)}} $ in a POMDP.

\ifco{
Each TMA is a graph of simpler local macro-actions (LMAs). Each LMA is a local feedback controller that maps current belief $ b^{(i)} $ of the $ i $-th agent to an action. A simple example is a linear LMA $ \mu(b)=-L(\hat{x}^{+}-\mathbf{v}) $, where $ L $ is the LMA gain, $ \hat{x}^{+} $ is the first moment (mean) of belief $ b $ and $ \mathbf{v} $ is the desired mean value. It can be shown that using appropriate filters to propagate the belief and appropriate LMA gains, the LMA can drive the system's belief to a particular region (attractor of LMA) in the belief space denoted by $ B=\{b:\|b-\check{b}\|\leq\epsilon \} $, where $ \check{b} $ is known for a given $ \mathbf{v} $ and gain $ L $ \cite{Ali14-IJRR}. We refer to $B$ as a milestone, which is comprised of a set of beliefs.

\vspace{-0pt}
\begin{algorithm}[t!]
\SetKwFunction{ConstructTMA}{ConstructTMA}
\caption{TMA Construction (Offline)}\label{alg:TMA-construction}
\textbf{Procedure}  : $\ConstructTMA(b,\mathbf{v}^{goal},\mathcal{M})$\\
\textbf{input}  :  Initial belief $b$, mean of goal belief $ \mathbf{v}^{goal} $, task POMDP $ \mathcal{M} $;\\
\textbf{output}  :  TMA policy $ \pi^* $, success probability of TMA $ P(B^{goal}|b_{0},\pi^*) $, value of taking TMA $ V(b_{0},\pi^*) $;\\
{
Sample a set of LMA parameters $ \{\theta^{j}\}_{j=1}^{n-1} $ from the state space of $ \mathcal{M} $, where $ \theta^{n-1} $ includes $ \mathbf{v}^{goal} $; \label{line:sample-theta}\\

Corresponding to each $ \theta^{j} $, construct a milestone $ B^{j} $ in belief space; \label{line:milestone}\\

Add to them the $ n $-th node as the singleton milestone $ B^{n}=\{b\} $;\\

Connect milestones using LMAs $ \mathcal{L}^{ij} $; \label{line:connect}\\

and compute the LMA rewards, execution time, and transition probabilities by simulating LMAs offline; \label{line:simulate}\\

Solve the LMA graph DP in \eqref{eq:FIRM-DP} to construct TMA $\pi^*$;\label{line:solve-DP}\\

Compute the associated success probability $ P(B^{goal}|b,\pi^*) $, completion time $ T(B^{j}|b,\pi^{*}) ~\forall j $, and value $ V(b,\pi^*)$;\label{line:compute-success-time}\\

\Return TMA policy $\pi^*$, success probability $ P(B^{goal}|b,\pi^*) $, completion time $ T(B^{j}|b,\pi^{*}),~\forall j $, and value $ V(b,\pi^*)$}
\end{algorithm}

Each TMA is constructed incrementally using sampling-based methods. Alg. \ref{alg:TMA-construction} recaps the construction of a TMA. We sample a set of LMA parameters $ \{\theta^{j} = (L^{j},\mathbf{v}^{j})\} $ (Line \ref{line:sample-theta}) and generate corresponding LMAs $ \{\mu^{j}\} $. Associated with the $ j $-th LMA, we compute the $ j $-th milestone $ \check{b}^{j} $ and its $\epsilon$-neighborhood milestone $ B^{j}=\{b:\|b-\check{b}^{j}\|\leq\epsilon \} $ (Line \ref{line:milestone}). We connect $ B^{j} $ to its $ k $-nearest neighbors via their corresponding LMAs. In other words, for neighboring nodes $ B^{i} $ and $ B^{j} $, we use $ \mathcal{L}^{ij}=\mu^{j} $ to take the belief from $ B^{i} $ to $ B^{j} $ (Line \ref{line:connect}). One can view a TMA as a graph whose nodes are $ \mathbb{V}=\{B^{j}\} $ and whose edges are LMAs  $\mathbb{L}=\{\mathcal{L}^{ij} \}$ (Fig. \ref{fig:hierarchy}). We denote the set of available LMAs at $ B^{i} $ by $ \mathbb{L}(i) $. To incorporate the lower-level state constraints (e.g., obstacles) and control constraints, we augment the set of nodes $ \mathbb{V} $ with a hypothetical node $ B^{0} $ \sXX{add in algorithm} that represents the constraints. Therefore, while taking any $ \mathcal{L}^{ij} $, there is a chance that system ends up in $ B^{0} $ (i.e., violates the constraints). For more details on this procedure see \cite{Shay14-Dec-POSMDP-TR,Ali14-IJRR}.
}

\iftr{
For any given agent, a feedback controller in partially-observable environment comprises a Bayesian filter $ b_{k+1}=\tau(b_{k},u_{k},z_{k+1}) $ that evolves the belief and a separated controller $ u_{k+1}=\mu(b_{k+1}) $ that generates control signals based on the current belief (figure). Therefore, a feedback controller $ \mathcal{L} $ in belief space can be viewed as a function that maps the current belief $ b_{k} $, control $ u_{k} $, and observation $ z_{k+1} $ to the pair of next belief $ b_{k+1} $ and control $ u_{k+1} $; i.e., $(b_{k+1},u_{k+1})=\mathcal{L}(b_{k},u_{k},z_{k+1})=(\tau(b_{k},u_{k},z_{k+1}), \mu(\tau(b_{k},u_{k},z_{k+1})))$.

A local MA we consider herein is a feedback controller that is effective (has basin of attraction) locally in a region of the state/belief space. Many controllers that rely on linearization fall into this category as the linearization assumption is valid in locally around the linearization point. The goal of LMA in the partially-observable setting is to drive the system's belief to a particular belief. In \cite{Ali14-IJRR} it has been shown that in Gaussian belief space, utilizing a combination of Kalman filter and linear controllers, the system's belief can be steered toward certain probability distributions. In other words, LMAs act like a funnel in belief space. Starting from a belief in the mouth of funnel (see Fig. \ref{fig:hierarchy}), the LMA drives the belief toward a target belief that is referred to as a \textit{milestone} herein.

Since LMAs act locally on belief space, we can locally linearize the system and design corresponding simple LMAs. In this paper, we assume the belief space is Gaussian and thus belief can be represented with a mean vector $\hat{x}^{+}$ and covariance matrix $ P^{+} $ denoted as $b\equiv(\hat{x}^{+},P^{+})$. For a given state point $ \mathbf{v} $, we linearize the nonlinear process and measurement equations to get a stationary linear system with Gaussian noises. Associated with this linear system, we design a stationary Kalman filter and a linear separated controller, $ \mu(b)=-L(\hat{x}^{+}-\mathbf{v}) $. Thus, the utilized LMA is parametrized by feedback gain matrix $ L $ and point $ \mathbf{v} $; i.e., $ \mu(b;\theta) $, where $ \theta=(L,\mathbf{v}) $. It can be shown that under appropriate choice of $ L $ and mild observability conditions, this Linear LMA acts as a funnel in belief space that drives the belief toward the milestone $ \check{b}\equiv(\check{x},\check{P}) $, where $ \check{x}=\mathbf{v} $ and $ \check{P} $ is the solution of the Riccati equation corresponding to the Kalman filter \cite{Ali14-IJRR}.

A chain of funnels is a sequence of funnels where the target belief of each funnel falls into the mouth (or pre-image) of the next funnel in the chain. A richer way of combining funnels is via graphication (to form a graph of funnels). An information roadmap is defined as a graph of funnels, where each node of this graph is a milestone and each edge is an LMA funnel.

To construct a graph of LMAs, we sample a set of parameters $ \{\theta^{j}\} $ and generate corresponding LMAs $ \{\mathcal{L}^{j}\} $. Associated with the $ j $-th LMA, we compute the $ j $-th milestone $ \check{b}^{j} $. We define the $ j $-th node of our LMA graph as an $ \epsilon $-neighborhood around the milestone; i.e., $ B^{j}=\{b~:~\|b-\check{b}^{j}\|\leq\epsilon \} $ and the set of all nodes as $ \mathbb{V}=\{B^{j}\} $. We connect $ B^{j} $ to its $ k $-nearest neighbors via their corresponding LMAs. If neighboring nodes $ i $ and $ j $ are so far from each other that the $ j $-th LMA $ \mathcal{L}^{j} $ cannot take the belief from $ i $ to $ j $ (since the linearization used to construct $ \mathcal{L}^{j} $ is not valid around $ B^{i} $), we utilize an edge controller (as detailed in \cite{Ali14-IJRR}). An edge controller is a finite-time controller whose role is to take the mean of distribution close enough to the node $ B^{j} $ via tracking a trajectory that connects $ \mathbf{v}^{i} $ to $ \mathbf{
v}^{j} $ in the state space. Once the distribution mean gets close enough to the target node, the system's control is handed over to the funnel associated with the target node. We denote the concatenation of the edge controller and the funnel utilized to take the belief from $ B^{i} $ to $ B^{j} $ by $ \mathcal{L}^{ij}$, which is defined as the the $ (i,j) $-th graph edge. The set of all edges are denoted by $\mathbb{L}=\{\mathcal{L}^{ij} \}$. We denote the set of available LMAs at $ B^{i} $ by $ \mathbb{L}(i) $. To incorporate the lower-level state constraints (e.g., obstacles) and control constraints, we consider $ B^{0} $ as a hypothetical node, hitting which represents violation of constraints. We add $ B^{0} $ to the set of nodes $ \mathbb{V} $. Therefore, taking any $ \mathcal{L}^{ij} $ there is a chance that system ends up in $ B^{0} $.
}

We can simulate the behavior of LMA $ \mathcal{L}^{ij} $ at $ B^{i} $ offline \ifco{(Line \ref{line:simulate})} and compute the probability of landing in any given node $ B^{r} $, which is denoted by $ P(B^{r}|B^{i},\mathcal{L}^{ij}) $. Similarly, we can compute the reward of taking LMA $ \mathcal{L}^{ij} $ at $ B^{i} $ offline, which is denoted by $ R(B^{i},\mathcal{L}^{ij}) $ and defined as the sum of one-step rewards under this LMA. Finally, by $ \mathcal{T}^{ij}=T(B^{i},\mathcal{L}^{ij}) $ we denote the time it takes for LMA $ \mathcal{L}^{ij} $ to complete its execution starting from $ B^{i} $.

\subsection{Utilizing TMAs in the Decentralized Setting} \label{sec:extendedTMAs}
In a decentralized setting, the following properties of the macro-action need to be available to the high-level decentralized planner: \textit{(i)} TMA value from any given initial belief, \textit{(ii)} TMA completion time from any given belief, and \textit{(iii)} TMA success probability from any given belief. What makes computing these properties challenging is the requirement that they need to be calculated for \textit{every} possible initial belief. Every belief is needed because when one agent's TMA terminates, the other agents might be in any belief while still continuing to execute their own TMA. This information about the progress of agents' TMAs is needed for nontrivial asynchronous TMA selection. 

In the following, we discuss how the graph-based structure of our proposed TMAs allows us to compute a closed-form equation for the success probability, value, and time. As a result, when evaluating the high-level decentralized policy, these values can be efficiently retrieved for any given start and goal states. This is particularly important in decentralized multi-agent planning since the state/belief of the $ j $-th agent is not known a priori when the MA of $ i $-th agent terminates.

A TMA policy is defined as a policy that is found by performing dynamic programming on the graph of LMAs. Consider a graph of LMAs that is constructed to perform a simple task such as open-the-door, pick-up-a-package, move-a-package, etc. \iftr{An important feature of this graph is that it is multi-query, meaning that it is valid for any starting and goal belief.} Depending on the goal belief of the task, we can solve the dynamic programming problem on the LMA graph that leads to a policy which achieves the goal while trying to maximize the accumulated reward and taking into account the probability of hitting failure set $ B^{0} $. Formally, we need to solve the following DP:
\begin{align}\label{eq:FIRM-DP}
&\!V^{*}(B^{i},\pi^{*}) \!= \!\!\!\max_{\mathcal{L} \in \mathbb{L}(i)} \Bigl( R(B^{i},\mathcal{L}) \!+\!\! \sum_{j}\!P(B^{j}| B^{i},\mathcal{L})V^{*}(B^{j}))\Bigr), \forall i\\
\nonumber&\pi^*(B^{i})\!= \!\arg\max_{\mathcal{L} \in \mathbb{L}(i)} \Bigl( R(B^{i},\mathcal{L}) \!+\!\! \sum_{j}P(B^{j}| B^{i},\mathcal{L})V^{*}(B^{j})\Bigr), \forall i
\end{align}
\ifco{where $ V^*(\cdot) $ is the optimal value defined over the graph nodes with $ V(B^{goal}) $ set to zero and $ V(B^{0}) $ set to a suitable negative reward for violating constraints.  Here, $\pi^*(\cdot) $ is the resulting TMA (Line \ref{line:solve-DP}). The primitive actions can be retrieved from the TMA via a two-stage computation: the TMA picks the best LMA at each milestone and the LMA generates the next action based on the current belief until the system reaches the next milestone; i.e., $ u_{k+1}=\pi^*(B)(b_{k})=\mathcal{L}(b_{k}) $ where $ B $ is the last visited milestone and $ \mathcal{L}=\pi^*(B) $ is the best LMA chosen by TMA at milestone $ B $. The space of TMAs is denoted as $ \mathbb{T}=\{\pi \} $.}
\iftr{where $ V^*(\cdot) $ is the optimal value defined over the graph nodes with $ V(B^{goal}) $ set to zero and $ V(B^{0}) $ set to a suitable negative reward for violating constraints. $ \pi^*(\cdot) $ is the resulting TMA. The primitive actions can be retrieved from TMA via a two-stage computation: TMA picks the best LMA at each milestone and LMA generates the next action based on the perceived observations until belief reaches the next milestone; i.e., $ u_{k+1}=\mathcal{L}(b_{k},u_{k},z_{k+1})=\pi^*(B)(b_{k},u_{k},z_{k+1}) $ where $ B $ is the last visited milestone and $ \mathcal{L}=\pi^*(B) $ is the best LMA chosen by TMA at milestone $ B $. The space of TMAs is denoted as $ \mathbb{T}=\{\pi \} $.}

For a given optimal TMA $ \pi^* $, the associated optimal value $ V^{*}(B^{i},\pi^*)$ from any node $ B^{i} $ is computed via solving $ \eqref{eq:FIRM-DP} $. Also, using Markov chain theory we can analytically compute the probability $ P(B^{goal}|B^{i},\pi^*) $ of reaching the goal node $ B^{goal} $ under the optimal TMA $ \pi^* $ starting from any node $ B^{i} $ in the offline phase \cite{Ali14-IJRR}. 


Similarly, we can compute the time it takes for the TMA to go from $ B^{i} $ to $ B^{goal} $ under $ \pi^{*} $ as follows:
\begin{align}\label{eq:time-dp}
\nonumber T^{g}(B^{i};\pi^{*}) &= T(B^{i},\pi^{*}(B^{i})) \\
 &+\sum_{j}P(B^{j}| B^{i},\pi^{*}(B^{i}))T^{g}(B^{j};\pi^{*}),~ \forall i
\end{align}
\ifco{\sXX{T, $\tau$ distinction}}
where $ T^{g}(B;\pi) $ denotes the time it takes for TMA $ \pi $ to take the system from $ B $ to TMA's goal. Defining $ \mathcal{T}^{i}=T(B^{i};\pi^{*}(B^{i})) $ and $ \bar{\mathcal{T}}=(\mathcal{T}^{1},\mathcal{T}^{2},\cdots,\mathcal{T}^{n})^{T} $ we can write (\ref{eq:time-dp}) in its matrix form as:
\begin{align}\label{eq:time-dp-matrix}
\bar{T}^{g} &= \bar{\mathcal{T}} + \bar{P}\bar{T}^{g}\Rightarrow
\bar{T}^{g}=(I-\bar{P})^{-1}\bar{\mathcal{T}}
\end{align}
where $ \bar{T}^{g} $ is a column vector with $ i $-th element equal to $ T^{g}(B^{i};\pi^{*}) $ and $ \bar{P} $ is a matrix with $ (i,j) $-th entry equal to $ P(B^{j}| B^{i},\pi^{*}(B^{i})) $.

Therefore, a TMA can be used in a higher-level planning algorithm as a MA whose success probability, execution time, and reward can be computed offline.

\subsection{Environmental State and Updated Model} \label{sec:e-state-dec-policy}
We also extend TMAs to the multi-agent setting where there is an environmental state that is locally observable by agents and can be affected by other agents.

We denote the environment state (e-state) at the $ k $-th time step as $x^{e}_{k} \in \mathbb{X}^{e} $. It encodes the information in the environment that can be manipulated and observed by different agents. We assume $x^{e}_{k}$ is only locally (partially) observable. 
An example for $ x^{e}_{k} $ in the package delivery application (presented in Section \ref{sec:experiments}) is ``there is a package in the base''. An agent can only get this measurement if the agent is in the base (hence it is partial). 

Any given TMA $ \pi $ is only available at a subset of e-states, denoted by $ \mathbb{X}^{e}(\pi) $. In many applications $ \mathbb{X}^{e}(\pi) $ is a small finite set. Thus, we can extend the cost and transition probabilities of TMA $ \pi $ for all $ x^{e}\in\mathbb{X}^{e} $ by performing the TMA evaluation described in Section \ref{sec:extendedTMAs} for all $ x^{e}\in\mathbb{X}^{e} $.

We extend transition probabilities $P(B^{goal}|b,\pi)$ to take the e-state into account, i.e., $ P(B^{goal},x^{e'}|b,x^{e},\pi) $, which denotes the probability of getting to the goal region $ B^{goal} $ and e-state $ x^{e'} $ starting from belief $ b $ and e-state $ x^{e} $ under the TMA policy $ \pi $. Similarly, the TMA's value function $ V(b,\pi) $ is extended to $ V(b,x^{e},\pi) $, for all $ x^{e}\in\mathbb{X}^{e}(\pi) $.

The joint reward $ \bar{R}(\bar{x},x^{e},\bar{u}) $ encodes the reward obtained by the entire team, where $ \bar{x}=(x^{(1)},\cdots,x^{(n)}) $ is the set of states for different agents and $ \bar{u}=(u^{(1)},\cdots,u^{(n)}) $ is the set of actions taken by all agents. 

We assume the joint reward is a multi-linear function of a set of reward functions $ \{R^{(1)},\cdots,R^{(n)}\} $ and $ R^{E} $, where $ R^{(i)} $ only depends on the $ i $-th agent's state and $ R^{E} $ depends on all the agents. In other words, we have:
\begin{align}
\nonumber\bar{R}(\bar{x},x^{e},\bar{u})&=g\left(R^{(1)}(x^{(1)},x^{e},u^{(1)}),
R^{(2)}(x^{(2)},x^{e},u^{(2)}),\right.\\
&\left.\cdots,R^{(n)}(x^{(n)},x^{e},u^{(n)}),R^{E}(\bar{x},x^{e},\bar{u})\right)
\end{align}
In multi-agent planning domains, often computing $ R^{E} $ is computationally less expensive than computing $ \bar{R} $, which is the property we exploit in designing the higher-level decentralized algorithm.
\ifco{\sXX{why?}.}

The joint reward $ \bar{R}(\bar{b},x^{e},\bar{u}) $ encodes the reward obtained by the entire team, where $ \bar{b}=(b^{(1)},\cdots,b^{(n)}) $ is the joint belief and $ \bar{u} $ is the joint action defined previously. 

Similarly, the joint policy $ \bar{\phi}=\{\phi^{(1)},\cdots,\phi^{(n)} \} $ is the set of all decentralized policies, where $ \phi^{(i)} $ is the decentralized policy associated with the $ i $-th agent. In the next section, we discuss how these decentralized policies can be computed based on the Dec-POSMDP formulation.

Joint value $ \bar{V}(\bar{b},x^{e},\bar{\phi}) $ encodes the value of executing the collection $ \bar{\phi} $ of decentralized policies starting from environment state $ x_{0}^{e} $ and initial joint belief $ \bar{b} $.
\section{The \ds/ Framework} \label{sec:decposmdp}
In this section, we formally introduce the  \ds/  framework. We discuss how the use of TMAs can transform a continuous Dec-POMDP to a Dec-POSMDP over finite number of MAs. This transformation allows discrete domain algorithms to generate a decentralized solution for general continuous problems. 

We denote the high-level decentralized policy for the $ i $-th agent by $ \phi^{(i)}:\Xi^{(i)}\rightarrow\mathbb{T}^{(i)} $, where $ \Xi^{(i)} $ is the macro-action history for the $ i $-th agent (as opposed to the action-observation history), which is formally defined as:
\begin{align}
\nonumber&\Xi^{(i)}_{k}\!\!=\!
(z^{e,(i)}_{1},\pi^{(i)}_{1}\!,H^{(i)}_{1}\!,z^{e,(i)}_{2}\!,\pi^{(i)}_{2}\!,H^{(i)}_{2}\!,\ldots,\pi^{(i)}_{n-1}\!,H^{(i)}_{n-1}\!,z^{e,(i)}_{n})
\end{align}
which includes the chosen macro-actions $ \{\pi^{(i)}_{k}\} $, the action-observation histories under chosen macro-actions $ \{H^{(i)}_{k}\} $, and the environmental observations $ \{z^{e,(i)}_{k}\} $ received at the termination of macro-actions. 
Accordingly, we can define a \emph{joint policy} $ \bar{\phi}=(\phi^{(1)},\phi^{(2)},\ldots,\phi^{(n)}) $ for all agents and a joint macro-action policy as $ \bar{\pi}=(\pi^{(1)},\pi^{(2)},\ldots,\pi^{(n)}) $.

Each time an agent completes a TMA, it receives an observation of the environmental state $ x^{e} $, denoted by $ o^{e} $. Also, due to the special structure of the proposed TMAs, we can record the agent's final belief $ b^{f} $ at TMA's termination (which compresses the entire history $ H $ under that TMA). We denote this pair as $ \breve{o}=(b^{f},o^{e}) $. As a result, we can compress the macro-action history as
$\Xi^{(i)}=\{\breve{o}^{(i)}_{1},\pi^{(i)}_{1},\breve{o}^{(i)}_{2},\pi^{(i)}_{2},\cdots,\breve{o}^{(i)}_{k-1},\pi^{(i)}_{k-1},\breve{o}^{(i)}_{k}\}$.

 
The value of joint policy $ \bar{\phi} $  is 
\begin{align}
\bar{V}^{\bar{\phi}}(\bar{b})=\mathbb{E} \left[ \sum_{t=0}^{\infty} \gamma^t\bar{R}( \bar{s}_{t},\bar{u}_{t}) | p(\bar{s}_{0})=\bar{b},\bar{\phi},\{\bar{\pi}\} \right],
\end{align}
but it is unclear how to evaluate this equation without a full (discrete) low-level model of the domain. Even in that case, it would often be intractable to calculate the value directly. 
Therefore, we will formally define the Dec-POSMDP problem, which has the same goal as the Dec-POMDP problem (finding the optimal policy), but in this case we seek the optimal policy for choosing macro-actions in our semi-Markov setting:
\begin{align}\label{eq:dec-posmdp}
\bar{\phi}^{*}=\arg\max_{\bar{\phi}}\bar{V}^{\bar{\phi}}
\end{align}

\begin{Def}
\textbf{(Dec-POSMDP)}
The \ds/ framework is described by the following elements
\end{Def}
\begin{itemize}
\item $\mathbb{I}=\{1,2,\cdots,n\}$ is a finite set of agents' indices.
\item $\mathbb{B}^{(1)}\times\mathbb{B}^{(2)}\times\ldots\times\mathbb{B}^{(n)}\times\mathbb{X}^{e} $ is the underlying state space for the proposed Dec-POSMDP, where $ \mathbb{B}^{(i)} $ is the set of beliefs of $ i $-th agent's TMAs (i.e., $ \mathbb{T}^{(i)} $).
\item $ \mathbb{T}=\mathbb{T}^{(1)}\times\mathbb{T}^{(2)}\ldots\times\mathbb{T}^{(n)} $ is the space of high-level actions in Dec-POSMDP, where $ \mathbb{T}^{(i)} $ is the set of TMAs for the $ i $-th agent.
\item $ P(\bar{b}',x^{e'},k|\bar{b},x^{e},\bar{\pi}) $ denotes the transition probability under TMAs $ \bar{\pi} $ from a given $ \bar{b},x^{e} $ to $ \bar{b}',x^{e'} $ as described below.
\item $ \bar{R}^{\tau}\!(\bar{b},x^{e},\bar{\pi}) $ denotes the reward/value of taking TMA $ \bar{\pi} $ at $ \bar{b},x^{e} $ as described below.
\item $ \breve{\mathbb{O}} $ is the set of environmental observations. 
\item $ P(\bar{\breve{o}}|\bar{b},x^{e}) $ denotes the observation likelihood model.
\end{itemize}
Again, a general Dec-POSDMP need not have a factored state space. Also, while the state space is very large, macro-actions allow much of it to be ignored once these high-level transition, reward and observation functions are calculated. 

Below, we describe the above elements in more details. For further explanation and derivations please see \cite{Shay14-Dec-POSMDP-TR}.
The planner $ \bar{\phi}=\{\phi^{(1)},\phi^{(2)},\cdots,\phi^{(n)}\} $ that we construct in this section is fully decentralized in the sense that each agent $ i $ has its own policy $\phi^{(i)}:\Xi^{(i)}\rightarrow\mathbb{T}^{(i)}$ that generates the next MA based on the history of MAs taken and the observation perceived solely by the $ i $-th agent. 

Each policy $ \phi^{(i)} $ is a discrete controller, represented by a (policy) graph \cite{JAAMAS09,Macdermed13}. Each node of the discrete controller corresponds to a MA. Note that different nodes in the controller could use the same MA. Each edge in this graph is an $ \breve{o} $. An example discrete controller for a package delivery domain is illustrated in Fig. \ref{fig:packageDeliveryPolicy}. 

Consider a set of MAs $\bar{\pi}=(\pi^{(1)}, \pi^{(2)},\cdots,\pi^{(n)})$ for the entire team. Incorporating terminal conditions for different agents, we can evaluate the set of decentralized MAs until at least one of them stops as: 
\begin{align}
\nonumber&\bar{R}^{\tau}\!(\bar{b},x^{e},\bar{\pi}) \!=\! \mathbb{E}\!\left[\sum_{t=0}^{\bar{\tau}_{min}}\!\!\gamma^{t}\bar{R}(\bar{x}_{k},x^{e}_{k},\bar{u}_{k})|\bar{\pi}, p(\bar{x}_{0})\!=\!\bar{b},x^{e}_{0}\!=\!x^{e}\right]
\end{align}
where \vspace{-20pt}
\begin{align}
\bar{\tau}_{min} = \min_{i}\min_{t}\{t:b^{(i)}_{t}\in B^{(i),goal} \}
\end{align}

It can be shown that the probability of transitioning between two configurations (from $\bar{b},x^{e} $ to $\bar{b}',x^{e'}$) after $ k $ steps under the set of MAs $ \bar{\pi} $ is given by:
\begin{align} 
\nonumber &P(\bar{b}',x^{e'},k|\bar{b},x^{e},\bar{\pi})=P(x^{e'}_k,\bar{b}'_k|x^{e}_{0},\bar{b}_{0},\bar{\pi})\\
\nonumber&=\sum_{x^{e}_{k-1},\bar{b}_{k-1}}\left[P(x^{e'}_k|x^{e}_{k-1},\bar{\pi}(\bar{b}_{k-1}))\times\right.\\
&\left.P(\bar{b}'_k|x^{e}_{k-1},\bar{b}_{k-1},\bar{\pi}(\bar{b}_{k-1}))P(x^{e}_{k-1},\bar{b}_{k-1}|x^{e}_{0},\bar{b}_{0},\bar{\pi})\right]
\end{align}

Joint value $ \bar{V}^{\bar{\phi}}(\bar{b},x^{e}) $ then encodes the value of executing the collection $ \bar{\phi} $ of decentralized policies starting from environment state $ x_{0}^{e} $ and initial joint belief $ \bar{b}_{0} $. The below equation describes the value transformation from the primitive actions to MAs, which is vital for allowing us to efficiently perform evaluation. Details of this derivation can be found in \cite{Shay14-Dec-POSMDP-TR}. 
\begin{align}\label{eqn:joint_value_eval}
\nonumber\bar{V}^{\bar{\phi}}(\bar{b}_{0},x^{e}_{0})&=\mathbb{E}\left[\sum_{t=0}^{\infty}\gamma^{t}\bar{R}(\bar{x}_{t},x^{e}_{t},\bar{u}_{t})|\bar{\phi}, \bar{b}_{0},x^{e}_{0}\right]\\
&=\mathbb{E}\left[\sum_{k=0}^{\infty}\gamma^{t_{k}}\bar{R}^{\tau}(\bar{b}_{t_{k}},x^{e}_{t_{k}},\bar{\pi}_{t_{k}})|\bar{\phi},\bar{b}_{0},x^{e}_{0}\right]
\end{align}
where $ t_{k}= \min_{i}\min_{t}\{t>t_{k-1}:b^{(i)}_{t}\in B^{(i),goal} \} $ and 
\begin{align}
\bar{\pi}=\bar{\phi}(\bar{b},x^{e}).
\end{align}

The dynamic programming formulation corresponding to the defined joint value function over MAs is:
\begin{align}
\nonumber\bar{V}^{\bar{\phi}}(\bar{b},&x^{e})= \bar{R}^{\tau}(\bar{b},x^{e},\bar{\pi})+\\
&\sum_{k=0}^\infty\gamma^k\sum_{\bar{b}',x^{e'}}P(\bar{b}',x^{e'},k|\bar{b},x^{e},\bar{\pi})\bar{V}^{\bar{\phi}}(\bar{b}',x^{e'})
\end{align}

The critical reduction from the continuous Dec-POMDP to the Dec-POSMDP over a finite number of macro-actions is a key factor in solving large Dec-POMDP problems. In the following section we discuss how we compute a decentralized policy based on the Dec-POSMDP formulation.

\subsection{Masked Monte Carlo Search (MMCS)}
In this section, we propose an efficient method, referred to as Masked Monte Carlo Search (MMCS), to generate a decentralized multi-agent solution to solve the Dec-POSMDP. 
As demonstrated in Section~\ref{sec:experiments}, MMCS allows extremely large problems to be solved. It uses an informed Monte Carlo policy sampling scheme to achieve this, exploiting results from previous policy evaluations to narrow the search space. 

\SetKwFunction{MMCS}{MMCS}
\begin{algorithm}[t]
\caption{MMCS}\label{alg:MMCS}
\textbf{Procedure}  : $\MMCS(\mathbb{T}, K_{d})$\\
\textbf{input}  : set of TMAs, $\mathbb{T}$, default number of best policies to check in each iteration, $K_{d}$\\
\textbf{output}  :  decentralized policy $ \bar{\phi} $\\
{
\ForEach{agent $i$}
 {
	$masked^{(i)} \gets$ setToFalse();\\
	$\phi^{(i)}_{d} \gets null $\label{line:mmcs_sampleValidPolicy};\\
 }
\For{$iter_{MMCS}=1$ to $iter_{max, MMCS}$} 
{
  \For{$iter_{MC}=1$ to $iter_{max, MC}$}
 {
  $\phi_{new} \gets \phi_{d}$;\\
  \ForEach{agent $i$}
  {
 
  \ForEach{$(\pi,\breve{o}) \in \mathbb{T} \times \breve{\mathbb{O}} $} 
  {
   	\If{not $masked^{(i)}(\pi,\breve{o})$}
   	{\label{line:mmcs_maskCheck}
   		$\phi^{(i)}_{new}(\pi,\breve{o}) \gets $sample$(\mathbb{T}(\pi,\breve{o}))$;\label{line:mmcs_sample}\\
   	}
  }
  }
  $\phi_{list}.$append$(\phi_{new})$;\\
  $\bar{V}^{\bar{\phi}}(\bar{b},x^{e}).$append$($evalPolicy$(\phi_{new}))$;\label{line:mmcs_evalPolicy}\\
}
 $K \gets K_{d}$;\\ 
 $\phi_{list} \gets $getBestKPolicies$(\phi_{list}, \bar{V}^{\bar{\phi}}(\bar{b},x^{e}), K)$;\\
 $(masked, \phi_{d}) \gets$ createMask($\phi_{list}$);\label{line:mmcs_createMask}\\
 $K \gets$ 1;\\
 $\bar{\phi} \gets $getBestKPolicies$(\phi_{list}, \bar{V}^{\bar{\phi}}(\bar{b},x^{e}), K)$;\label{line:mmcs_chooseBestPolicy}\\
 
}
\Return $\bar{\phi}$;\\
}
\end{algorithm}

Because the infinite-horizon problem is undecidable \cite{Madani99}, infinite-horizon methods typically  focus on producing approximate solutions given a computational budget \cite{Macdermed13,JAAMAS09}. A Monte Carlo approach can be implemented by repeatedly randomly sampling from the policy space and retaining the policy with the highest expected value as an approximate solution. The search can be stopped at any point and the latest iteration of the approximate solution can be utilized. 

The MMCS algorithm is detailed in Alg. \ref{alg:MMCS}. 
MMCS uses a discrete controller to represent the policy, $\phi^{(i)}$, of each agent. The nodes of the discrete controller are TMAs, $\pi \in \mathbb{T}$, and edges are observations, $\breve{o} \in \breve{\mathbb{O}}$. Each agent $i$ transitions in the discrete controller by completing a TMA $\pi_{k}$, seeing an observation $\breve{o}_{k}$, and following the appropriate edge to the next TMA, $\pi_{k+1}$. Fig. \ref{fig:packageDeliveryPolicy} shows part of a single robot's policy.

MMCS initially generates a set of valid policies by randomly sampling the policy space (Line \ref{line:mmcs_sample}), while adhering to the constraint that the termination set of a TMA node in the discrete controller must intersect the initiation set of its child TMA node. The joint value, $ \bar{V}^{\bar{\phi}}(\bar{b},x^{e}) $, of each policy given an initial joint belief, $\bar{b}$, and e-state, $x^{e}$, is calculated by repeatedly applying the policy (either in a simulator or a real-world system) and taking its expected value (Line \ref{line:mmcs_evalPolicy}).

MMCS identifies the TMA transitions that occur most often in the set of policies with the $K$ highest joint values, in a process called `masking' (Line \ref{line:mmcs_createMask}). It then opts to include these transitions in future iterations of the policy search by explicitly checking for the existence of a mask for that transition, preventing the transition from being re-sampled (Line \ref{line:mmcs_maskCheck}). Note that the mask is not permanent, as re-evaluation of the mask using the best $K$ policies occurs in each iteration of MMCS (Line \ref{line:mmcs_createMask}). 

The above process is repeated until a computational budget is reached, after which the best policy is selected as the approximate solution to the Dec-POSMDP (Line \ref{line:mmcs_chooseBestPolicy}).
Our algorithm offers the advantage of balancing exploration and exploitation of the search space. Although `masking' places focus on promising policies, it is done in conjunction with random sampling, allowing previously unexplored regions of the policy space to be sampled.
\ifco{\cXX{Say something about convergence here? With random sampling, we can provide a probabilistic guarantee on the solution quality of the controllers that are generated. Specifically, we can show that with probability at least $1-\delta$ that we can construct controllers that have value within $\epsilon$ of the optimal value \cite{AAMAS09}. For general infinite-horizon problems this would require much larger controllers and would be completely infeasible to generate any reasonable bound, but we don't need to mention that :)}}

\section{Experiments} \label{sec:experiments}
In this section, we consider a package delivery under uncertainty scenario involving a team of heterogeneous robots, an application which has recently received particular attention \cite{Ali14-package-IROS,Amazon13_Forbes}. The overall objective in this problem is to retrieve and deliver packages from base locations to delivery locations using a group of robots. 

Though our method is general and can be used for many decentralized planning problems, this domain was chosen due to its extreme complexity. For instance, the policy space of the posed problem has cardinality $5.622e+17$, \ifco{\cXX{The policy space is technically infinite (since controllers could be any size). Do you mean when fixing the controller size? If so, you should say what you are fixing the controller size to.}} making it computationally intractable to solve. Additional challenges stem from the presence of different sources of uncertainty (wind, actuator, sensor), obstacles and constraints in the environment, and different types of tasks (pick-up, drop-off, etc.). Also, the presence of joint tasks such as joint pickup of large packages introduces a significant multi-robot coordination component. This problem is formulated as a Dec-POMDP in continuous space, and hence current Dec-POMDP methods are not applicable. 
Even if we could modify the domain to allow them to be used, simply discretizing the state/action/observation space in this problem would lead to poor solution quality or a computationally intractable Dec-POMDP.

Fig.~\ref{fig:Experiment Scenario} illustrates the package delivery domain. Robots are classified into two categories: air vehicles (quadcopters) and ground vehicles (trucks). Air vehicles handle pickup of packages from bases, and can also deliver packages to two delivery locations, $Dest_1$ and $Dest_2$. An additional delivery location $Dest_r$ exists in a regulated airspace, where air vehicles cannot fly. Instead, packages destined for the regulated airspace zone must be handed off to a ground vehicle at a rendezvous location. The ground vehicle is solely responsible for deliveries in this regulated region. Rewards are given to the team only when a package is dropped off at its correct delivery destination.

Packages are available for pickup at two bases in the domain. Each base contains a maximum of one package, and a stochastic generative model is used for allocating packages to bases. Each package has a designated delivery location, $\delta \in \Delta = \left\lbrace d_1, d_2, d_r \right\rbrace$. Additionally, each base has a size descriptor for its package, $\psi \in \Psi = \left\lbrace \varnothing, 1, 2 \right\rbrace$, where $\psi = \varnothing$ indicates no package at the base, $\psi = 1$ indicates a small package, and $\psi = 2$ indicates a large package. Small packages can be picked up by a single air vehicle, whereas large packages require cooperative pickup by two air vehicles. The descriptors of package destinations and sizes will implicitly impact the policy of the decentralized planner.

To allow coordination of cooperative TMAs, an environment variable stating the availability of a nearby vehicle, $\phi \in \Phi = \left\lbrace 0, 1 \right\rbrace $, is observable by robots at any base or at the rendezvous location, where $\phi = 1$ signifies that another robot is at the same milestone (or location) as the current robot and $\phi = 0$ signifies that all other robots are outside of some radius of the current robot's location. 

\begin{figure}[t!]
  \centering
 \includegraphics[width=0.35\textwidth]{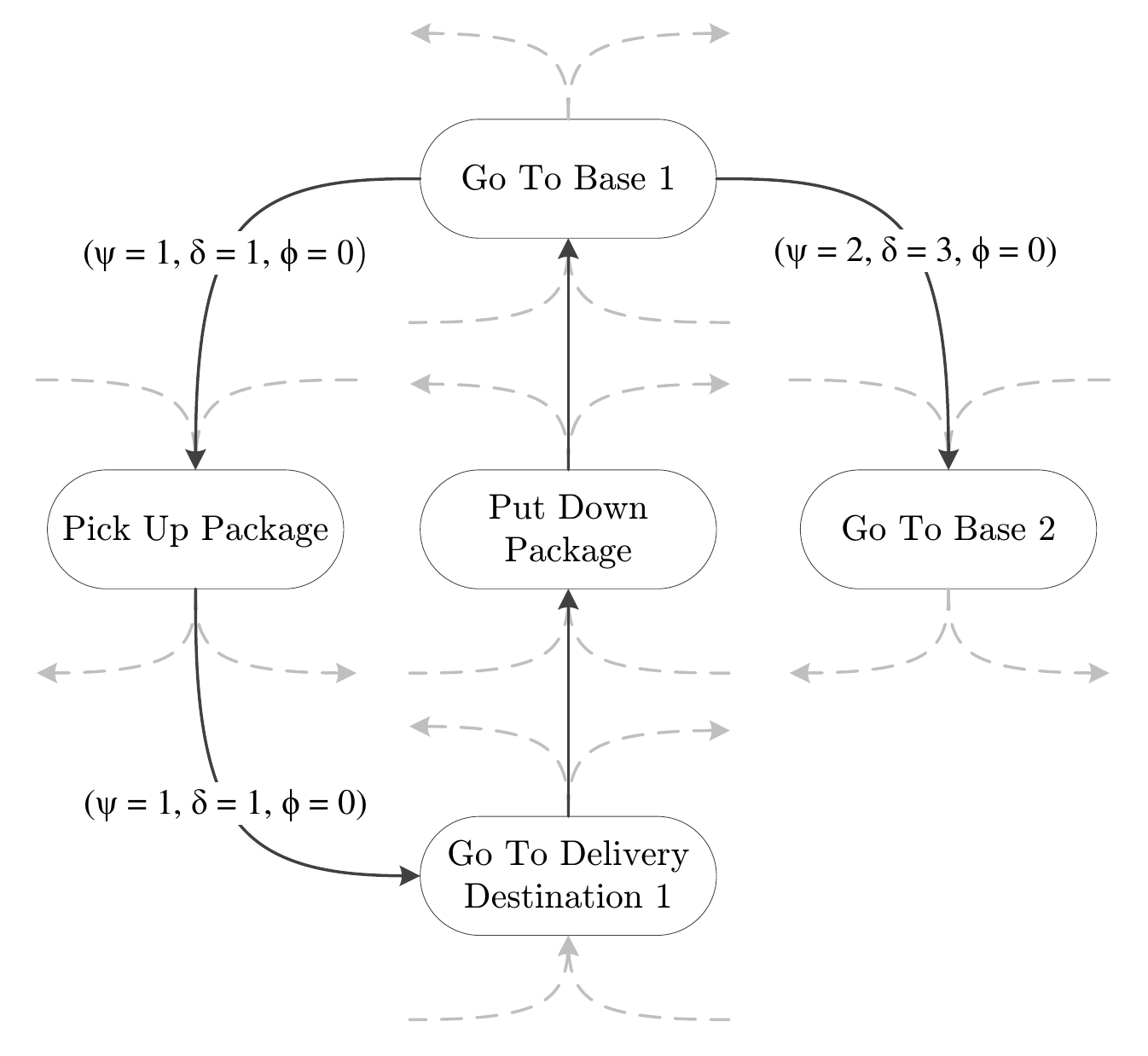}
  \vspace{-10pt}
 \caption{Partial segment of a single robot's policy obtained using MMCS for the package delivery domain. In this discrete policy controller, nodes represent TMAs and edges represent e-states. Greyed out edges represent connections to additional nodes which have been excluded to simplify the figure.}
 \label{fig:packageDeliveryPolicy}
\end{figure}


An robot at any base location can, therefore, observe the environmental state $x^{e} = ( \psi, \delta, \phi ) \in \Psi \times \Delta \times \Phi$, which contains details about the availability and the size of the package at the base (if it exists), the delivery destination of the package, and availability of nearby robots (for performing cooperative tasks). An robot at the rendezvous location can observe $x^{e} = \phi \in \Phi$ for guidance of rendezvous TMAs.

In this mission, we assume one ground robot and two air robots are used. We consider two base locations $ Base_{1} $ and $ Base_{2} $. Air robots are initially located at $Base_{1}$ and the ground robot is at $ Dest_{r} $. If $ b^{i}\in Base_{j} $, we say the $ i $-th robot is at the $j$-th base. 


The available TMAs in this domain are:
\begin{itemize}
\item Go to base $Base_{j}$ for $j \in \{1,2\} $
\iftr{\begin{itemize}
\item \emph{robots involved:} 1 air robot.
\item \emph{Initiation set:} Air robot $i$ is available, where $i \in \left\lbrace i_{a,1}, i_{a,2} \right\rbrace$. \ifco{\cXX{What does `robot $i$ is available' mean formally? It is in some belief set? }}
\item \emph{Termination set:} robot $i$ available and its belief is $b^i \in B^{h_j}$. 
\end{itemize}}

\item Go to delivery destination $Dest_{j}$ for $j \in \{1,2,r\} $
\iftr{
\begin{itemize}
\item \emph{robots involved:} 1 (any type)
\item \emph{Initiation set:} robot $i$ available, can be at any location.
\item \emph{Termination set:} robot $i$ available and its belief is $b^i  \in B^{d_j}$.
\end{itemize}
}

\item Joint go to delivery destination $Dest_{j}$ for $j \in \{1,2\} $
\iftr{
\begin{itemize}
\item \emph{robots involved:} 2 air robots.
\item \emph{Initiation set:}  Air robots $i_{a,1}$ and $i_{a,2}$ are available. 
\item \emph{Termination set:} robots $i_{a,1}$ and $i_{a,2}$ are available and their beliefs are $b^{i_{a,1}}  \in B^{d_j}$, $b^{i_{a,2}} \in B^{d_j}$, where $j \in \left\lbrace 1,2 \right\rbrace$.
\end{itemize}
}

\item Pick up package
\iftr{
\begin{itemize}
\item \emph{robots involved:} 1 air robot
\item \emph{Initiation set:} Air robot $i$ is available, where $i \in \left\lbrace I_{a,1},I_{a,2} \right\rbrace$. $x^e = \left\lbrace \psi = 1, \delta \in \Delta, \phi \in \Phi \right\rbrace$
\item \emph{Termination set:} Ground robot $i$ is carrying package and is unavailable. 
\end{itemize}
}

\item Joint pick up package
\iftr{
\begin{itemize}
\item \emph{robots involved:} 2 air robots.
\item \emph{Initiation set:} Air robots $i_{a,1}$ and $i_{a,2}$ are available. $x^e = \left\lbrace \psi = 2, \delta \in \Delta, \phi = exact \right\rbrace$
\item \emph{Termination set:} robots $i_{a,1}$ and $i_{a,2}$ are carrying package and are unavailable. 
\end{itemize}
}

\item Put down package
\iftr{
\begin{itemize}
\item \emph{robots involved:} 1 ground robot or 1 air robot.
\item \emph{Initiation set:} robot $i \in I$ is carrying package, $b^i \in B^{d_j}$, where $j \in \left\lbrace 1,2 \right\rbrace$
\item \emph{Termination set:} robot $i$ is available.
\end{itemize}
}

\item Joint put down package
\iftr{
\begin{itemize}
\item \emph{robots involved:} 2 air robots.
\item \emph{Initiation set:} Air robots $i_{a,1}$ and $i_{a,2}$ are available and jointly carrying a package. $b^{i_1} \in B^{d_j}$ and $b^{i_2} \in B^{d_j}$, where $j \in \left\lbrace 1,2 \right\rbrace$.
\item \emph{Termination set:} robots $i_{a,1}$ and $i_{a,2}$ are available and their beliefs are $b^{i_{a,1}} \in B^{d_j}$, $b^{i_{a,2}} \in B^{d_j}$, where $j \in \left\lbrace 1,2 \right\rbrace$.
\end{itemize}
}

\item Go to rendezvous location
\iftr{
\begin{itemize}
\item \emph{robots involved:} 1 ground robot or 1 air robot
\item \emph{Initiation set:} robot $i \in I$ is available.
\item \emph{Termination set:} robot $i$ is available and its belief is $b^i \in B^{d_r}$.
\end{itemize}
}

\item Place package on truck
\iftr{
\begin{itemize}
\item \emph{robots involved:} 1 air robot, 1 ground robot.
\item \emph{Initiation set:} Air robot $i_a \in \left\lbrace I_{a,1},I_{a,2} \right\rbrace$ and ground robot $i_g$ are available and located at rendezvous location, where $b^{i_a} \in B^{d_r}$ and $b^{i_a} \in B^{d_r}$.
\item \emph{Termination set:} robots $i_a$ and $i_g$ are available and their beliefs are $b^{i_a}  \in B^{d_j}$, $b^{i_g}  \in B^{d_j}$, where $j \in \left\lbrace 1,2 \right\rbrace$.
\end{itemize}
}

\item Wait at current location
\iftr{
\begin{itemize}
\item \emph{robots involved:} 1 ground robot or 1 air robot.
\item \emph{Initiation set:} robot $i \in I$ is available.
\item \emph{Termination set:} robot $i$ is available.
\end{itemize}
}
\end{itemize}

Each TMA is defined with an associated initiation and termination set. For instance, ``Go to $Base_{j}$" TMA is defined:
\begin{itemize}
\item \emph{robots involved:} 1 air robot.
\item \emph{Initiation set:} An air robot $i$ is available. 
 \ifco{\cXX{What does `robot $i$ is available' mean formally? It is in some belief set? }}
\item \emph{Termination set:} robot $i$ available and $b^{(i)} \in Base_{j}$.
\end{itemize}
The robot's belief and the environmental state affect the TMA's behavior when undertaken. For instance, the behavior of the ``Go to $Base_{j}$" TMA will be affected by the presence of obstacles for different robots. For details on the definition of other TMAs,  see \cite{Shay14-Dec-POSMDP-TR}.

To generate a closed-loop policy corresponding to each TMA, we follow the procedure explained in Section \ref{sec:FIRM}. For example, for the ``Go to $Base_j$'' TMA, the state of system consists of the air robot's pose (position and orientation). Thus, the belief space for this TMA is the space of all possible probability distributions over the system's pose. To follow the procedure in Section \ref{sec:FIRM}, we incrementally sample beliefs in the belief space, design corresponding LMAs, generate a graph of LMAs, and solve dynamic programming on this graph to generate the TMA policy. Fig. \ref{fig:Value} shows the performance of an example TMA (``Go to $ Dest_{1} $''). The results show that the macro-action is optimized to achieve the goal in the manner that produces the highest value, but the performance is robust to noise and tends to minimize the constraint violation probability. The most important observation is that the value function is available over the entire space. The same is true for the success probability and completion time of the TMA. This information can then be directly used by the MMCS algorithm to perform evaluation of policies that use this macro-action.

\begin{figure}[b!]
  \centering
 \includegraphics[width=0.3\textwidth]{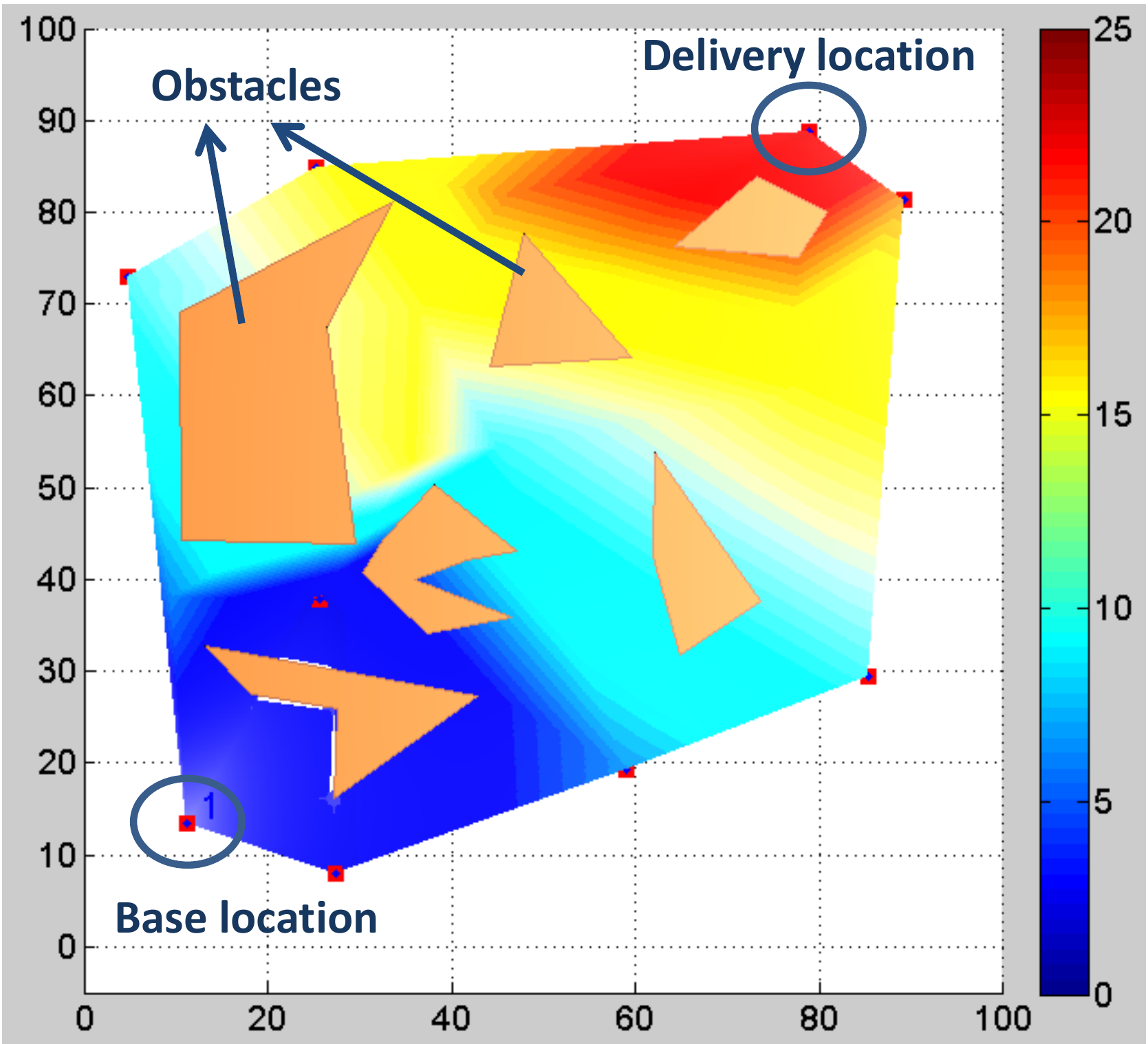}
  \vspace{-5pt}
 \caption{This figure shows the value of ``Go to $ Dest_{1} $'' TMA over its belief space (only the 2D mean part is shown).}
 \label{fig:Value}
\end{figure}

A portion of an example policy for a single air robot in the package delivery domain is illustrated in Fig.~\ref{fig:packageDeliveryPolicy}. The policy involves going to $Base_1$ and observing $x^e$. Subsequently, the robot chooses to either pick up the package (alone or with another robot) or go to $Base_2$. The full policy controller includes more nodes than shown in Fig. \ref{fig:packageDeliveryPolicy} (for all possible TMAs) as well as edges (for all possible environment observations). 

\begin{figure}[t]
  \centering
 \includegraphics[width=0.375\textwidth]{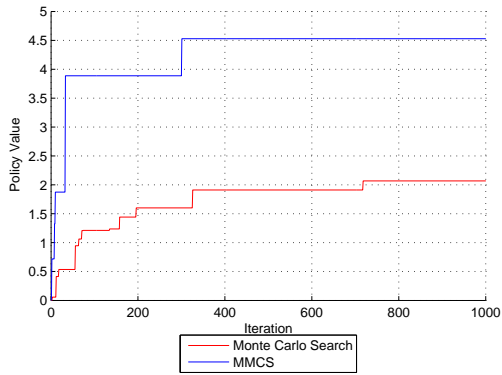}
 \vspace{-5pt}
 \caption{Comparison of best policy value over 1000 iterations for Monte Carlo and MMCS.}
 \label{fig:comparisonBestValues}
\end{figure}

\begin{figure}[t]
  \centering
 \includegraphics[width=0.35\textwidth]{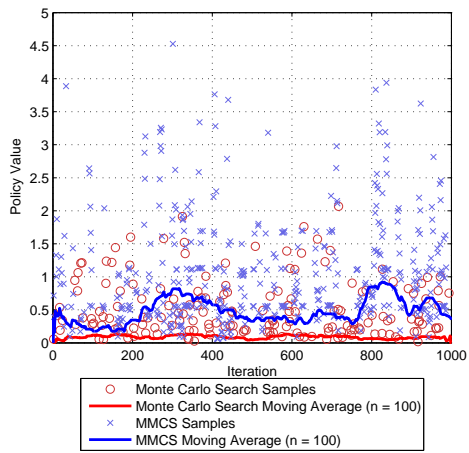}
 \caption{Comparison of policy sampling differences between Monte Carlo and MMCS. MMCS allows exploration of the policy space while exploiting knowledge of promising policies.}
  \vspace{-3pt}
 \label{fig:comparisonAllValues}
\end{figure}

\begin{figure}[t]
	\centering
	\includegraphics[width=0.35\textwidth]{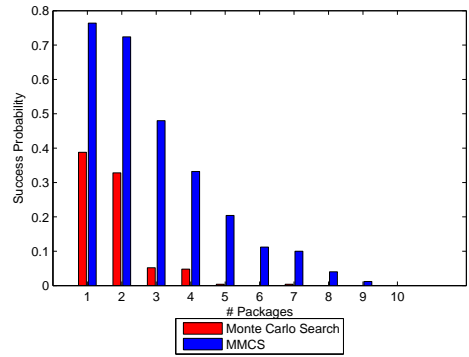}
	\caption{Success probability of delivering different numbers of packages with a fixed time horizon. Note that this is probability of delivering the specified number of packages \emph{or more}.}
	\vspace{-3pt}
	\label{fig:policySuccessProb}
\end{figure}

\begin{table}[t]
\caption{Comparison of search algorithms.}
\begin{center}
\begin{tabular}{|c||c||c|}
\hline
\bfseries Algorithm & \bfseries Policy Value & \bfseries Policy Iterations\\
\hline
Monte Carlo Search & 2.068 & 1000\\
\hline
MMCS & 4.528 & 1000\\
\hline 
Exhaustive Search & --- & 5.622e+17\\
\hline
\end{tabular}
\end{center}
\label{table:comparisonBestValues}
 \vspace{-15pt}
\end{table}

Fig. \ref{fig:comparisonBestValues} compares a uniform random Monte Carlo search to MMCS in the package delivery domain. Results for the Monte Carlo search were generated by repeatedly sampling random, valid policies and retaining the policy with the highest expected value. Both approaches used a policy controller with a fixed number of nodes ($n_{nodes} = 13$). 

Results indicate that given a fixed computational budget (quantified by the number of search iterations), MMCS outperforms standard Monte Carlo search in terms of expected value. Specifically, as seen in Table \ref{table:comparisonBestValues}, after 1000 search iterations in the package delivery problem, the expected value of the policy from MMCS is 118\% higher than the one obtained by Monte Carlo search. Additionally, due to this problem's extremely large policy space, determination of an optimal joint value through exhaustive search is not possible.

To more intuitively quantify performance of MMCS and Monte Carlo policies in the package delivery domain, Fig. \ref{fig:policySuccessProb} compares success probability of delivering a given minimum number of packages for both methods, within a fixed mission time horizon. Results were generated over 250 simulated runs with randomly-generated package types and initial conditions for robots. Using the Monte Carlo policy, probability of successfully delivering more than 2 packages given a fixed time horizon is almost negligible, whereas the policy resulting from MMCS successfully delivered a larger number of packages (in some cases up to 9). 

The experiments indicate that the Dec-POSMDP framework and our MMCS algorithm allow high-quality solutions to be generated for 
complex problems, such as multi-robot package delivery, that would not be possible using traditional Dec-POMDP approaches. 
\ifco{\cXX{We don't even talk about Figures 4 and 6? }}

\section{Conclusion} \label{sec:conclusion}
This paper proposed a layered framework to exploit the macro-action abstraction in decentralized planning for a group of robots acting under uncertainty. It formally reduces the Dec-POMDP problem to a Dec-POSMDP that can be solved using discrete space search techniques. 
We have additionally formulated an algorithm, MMCS, for solving Dec-POSMDPs and have shown its performance on a multi-robot package delivery problem under uncertainty. 
The Dec-POSMDP represents a general formalism for probabilistic multi-robot coordination problems and our results show that high-quality solutions can be automatically generated from this high-level domain specification. 
 \linebreak

 

\bibliographystyle{plain}
\bibliography{TemplateFiles/AliAgha}

\begin{thebibliography}{10}

\bibitem{Ali14-IJRR}
Ali-akbar Agha-mohammadi, Suman Chakravorty, and Nancy Amato.
\newblock {FIRM}: Sampling-based feedback motion planning under motion
  uncertainty and imperfect measurements.
\newblock {\em International Journal of Robotics Research (IJRR)},
  33(2):268--304, 2014.

\bibitem{Ali14-package-IROS}
Ali-akbar Agha-mohammadi, N.~Kemal Ure, Jonathan~P. How, and John Vian.
\newblock Health aware stochastic planning for persistent package delivery
  missions using quadrotors.
\newblock In {\em International Conference on Intelligent Robots and Systems
  (IROS)}, Chicago, September 2014.

\bibitem{JAAMAS09}
Christopher Amato, Daniel~S. Bernstein, and Shlomo Zilberstein.
\newblock Optimizing fixed-size stochastic controllers for {POMDPs} and
  decentralized {POMDPs}.
\newblock {\em Journal of Autonomous Agents and Multi-Agent Systems},
  21(3):293--320, 2010.

\bibitem{ICRA15MacDec}
Christopher Amato, George~D. Konidaris, Gabriel Cruz, Christopher~A. Maynor,
  Jonathan~P. How, and Leslie~P. Kaelbling.
\newblock Planning for decentralized control of multiple robots under
  uncertainty.
\newblock In {\em IEEE International Conference on Robotics and Automation
  (ICRA)}, 2015.

\bibitem{AAMAS14AKK}
Christopher Amato, George~D. Konidaris, and Leslie~P. Kaelbling.
\newblock Planning with macro-actions in decentralized {POMDPs}.
\newblock In {\em International Conference on Autonomous Agents and Multiagent
  Systems}, 2014.

\bibitem{AAMAS09}
Christopher Amato and Shlomo Zilberstein.
\newblock Achieving goals in decentralized {POMDPs}.
\newblock In {\em International Conference on Autonomous Agents and Multiagent
  Systems}, pages 593--600, 2009.

\bibitem{Amazon13_Forbes}
Steve Banker.
\newblock Amazon and drones -- here is why it will work, December 2013.

\bibitem{Bernstein02}
Daniel~S. Bernstein, Robert Givan, Neil Immerman, and Shlomo Zilberstein.
\newblock The complexity of decentralized control of {M}arkov decision
  processes.
\newblock {\em Mathematics of Operations Research}, 27(4):819--840, 2002.

\bibitem{Bernstein01}
Daniel~S. Bernstein, Shlomo Zilberstein, Richard Washington, and John~L.
  Bresina.
\newblock Planetary rover control as a {M}arkov decision process.
\newblock In {\em Proceedings of the the International Symposium on Artificial
  Intelligence, Robotics and Automation in Space}, 2001.

\bibitem{Emery05}
Rosemary Emery-Montemerlo, Geoff Gordon, Jeff Schneider, and Sebastian Thrun.
\newblock Game theoretic control for robot teams.
\newblock In {\em IEEE International Conference on Robotics and Automation
  (ICRA)}, pages 1163--1169, 2005.

\bibitem{He11JAIR}
Ruijie He, Emma Brunskill, and Nicholas Roy.
\newblock Efficient planning under uncertainty with macro-actions.
\newblock {\em Journal of Artificial Intelligence Research}, 40:523--570,
  February 2011.

\bibitem{Kober13}
Jens Kober, J.~Andrew Bagnell, and Jan Peters.
\newblock Reinforcement learning in robotics: A survey.
\newblock {\em The International Journal of Robotics Research}, 32(11):1238 --
  1274, September 2013.

\bibitem{Kumar-book-86}
P.~R. Kumar and P.~P. Varaiya.
\newblock {\em Stochastic Systems: Estimation, Identification, and Adaptive
  Control}.
\newblock Prentice-Hall, Englewood Cliffs, NJ, 1986.

\bibitem{Macdermed13}
Liam~C MacDermed and Charles Isbell.
\newblock Point based value iteration with optimal belief compression for
  {Dec-POMDPs}.
\newblock In {\em Advances in Neural Information Processing Systems}, pages
  100--108, 2013.

\bibitem{Madani99}
Omid Madani, Steve Hanks, and Anne Condon.
\newblock On the undecidability of probabilistic planning and infinite-horizon
  partially observable {M}arkov decision problems.
\newblock In {\em Proceedings of the Sixteen Conference on Artificial
  Intelligence (AAAI)}, pages 541--548, 1999.

\bibitem{Shay14-Dec-POSMDP-TR}
Shayegan Omidshafiei, Ali akbar Agha-mohammadi, Christopher Amato, and
  Jonathan~P. How.
\newblock Technical report: Decentralized control of partially observable
  markov decision processes using belief space macro-actions.
\newblock Technical report, Department of Aeronautics and Astronautics,
  Massachusetts Institute of Technology, September 2014.

\bibitem{Sutton99}
Richard~S Sutton, Doina Precup, and Satinder Singh.
\newblock Between {MDPs} and semi-{MDPs}: A framework for temporal abstraction
  in reinforcement learning.
\newblock {\em Artificial Intelligence}, 112(1):181--211, 1999.

\bibitem{Theocharous04}
Georgios Theocharous and Leslie~Pack Kaelbling.
\newblock Approximate planning in {POMDPs} with macro-actions.
\newblock In {\em Advances in Neural Information Processing Systems 16
  (NIPS03)}, 2004.

\bibitem{Ure13_BookChapter}
Nazim~Kemal Ure, Girish Chowdhary, Jonathan~P How, and John Vian.
\newblock {\em {Planning Under Uncertainty}}, chapter {Multi-Agent Planning for
  Persistent Surveillance}.
\newblock MIT Press, 2013.

\bibitem{Winstein13}
Keith Winstein and Hari Balakrishnan.
\newblock {TCP ex Machina: Computer-Generated Congestion Control}.
\newblock In {\em SIGCOMM}, 2013.

\end{thebibliography}

\end{document}